\documentclass[prl,twocolumn,superscriptaddress,showpacs,amsmath,amssymb]{revtex4-1}
\usepackage[dvips]{graphicx}
\usepackage[dvips]{color}
\usepackage{amsmath,subfigure,psfrag}
\usepackage{color}
\usepackage{natbib}
\usepackage{bm}
\def\U#1{{\rm #1}} 
\def\u#1{_{\rm #1}}
\newcommand{\ket}[1]{| #1 \rangle}
\newcommand{\bra}[1]{\langle #1 |}
\newcommand{\ketbra}[2]{| #1 \rangle \langle #2 |}
 
\def\H{{\rm H}}
\def\V{{\rm V}}
\def\HH{\U{HH}}
\def\VV{\U{VV}}

\def\D{\U{D}}
\def\A{\U{A}}
\def\Abar{\overline{\U{A}}}
\def\S{\U{S}}
\def\R{\U{R}}

\def\Mf{M\u{f}}
\def\Mb{M\u{b}}
\def\Nf{N\u{f}}
\def\Nb{N\u{b}}

\begin{document}
\title{
Experimental demonstration of robust entanglement distribution 
over reciprocal noisy channels assisted by a counter-propagating classical reference light
}
\author{Rikizo Ikuta}
\affiliation{Graduate School of Engineering Science, Osaka University,
Toyonaka, Osaka 560-8531, Japan}
\author{Shota Nozaki}
\affiliation{Graduate School of Engineering Science, Osaka University,
Toyonaka, Osaka 560-8531, Japan}
\author{Takashi Yamamoto}
\affiliation{Graduate School of Engineering Science, Osaka University,
Toyonaka, Osaka 560-8531, Japan}
\author{Masato Koashi}
\affiliation{Photon Science Center, 
The University of Tokyo, Bunkyo-ku, Tokyo 113-8656, Japan}
\author{Nobuyuki Imoto}
\affiliation{Graduate School of Engineering Science, Osaka University,
Toyonaka, Osaka 560-8531, Japan}

\begin{abstract}
 We experimentally demonstrate 
 a proposal~[Phys. Rev. A {\bf 87}, 052325~(2013)] of a scheme 
 for robust distribution of polarization entangled photon pairs 
 over collective noisy channels having the reciprocity. 
 Although the scheme employs the robustness of two qubit decoherence-free subspace, 
 by utilizing the forward propagation of one half of the entangled photons 
 and the backward propagation of a classical reference light, 
 it achieves an entanglement-sharing rate 
 proportional to the transmittance of the quantum channel for the signal photon. 
 We experimentally observed the efficient sharing rate 
 while keeping a highly entangled state after the transmission. 
 We also show that the protection method is applicable 
 to transmission of arbitrary polarization state of a single photon. 
\end{abstract}
\maketitle

Faithful and efficient distribution 
of photonic entangled states through noisy and lossy quantum channels 
is important for realizing various kinds of 
quantum information processing, 
such as quantum key distribution~\cite{1984Bennett, 1991Ekert, 1992Bennett}, 
quantum repeaters~\cite{2011Sangouard}, 
and quantum computation between distant parties~\cite{2009Broadbent, 2016Takeuchi}. 
A decoherence-free subspace~(DFS) formed by multiple qubits 
is useful to overcome fluctuations during the transmission 
which cause disturbance on quantum states. 
So far, a lot of proposals and demonstrations
for faithful transmission of photonic quantum states 
in a DFS against collective noises have been
actively studied~\cite{2000Kwiat,2003Walton,2004Boileau,
2004Bourennane,2004Boileau2,2005Yamamoto,2006Chen,
2007Prevedel,2007Yamamoto,2008Yamamoto}. 
However, for DFS protocols formed by two or more photons to succeed, 
all of the photons must arrive at the receiver side, 
which seriously limits the distribution efficiency of quantum states. 
When a two-photon DFS is used for faithful quantum communication 
over a dephasing channel~\cite{2003Lidar, 2005Yamamoto}, 
the transmission rate of the quantum state is proportional to $T^2$, 
where $T$ is the transmittance of a single photon. 
When we consider a random unitary~(depolarizing) quantum channel, 
a four-photon DFS is 
needed to encode a signal photon state~\cite{2001Kempe,2003Lidar, 2004Bourennane}, 
which leads to a transmission rate in the order of $T^4$. 

The inefficiency of such early DFS schemes 
has been resolved in the case of entangled photon pairs distributed 
over the dephasing channel~\cite{2011Ikuta}. 
The scaling of the achieved efficiency of sharing entanglement is proportional 
to $T$ instead of $T^2$. 
The key idea to improve the efficiency in the scheme 
is to prepare a reference single photon for the DFS from a coherent light pulse 
with average photon number of $\mathcal{O}(T^{-1})$ 
which backward-propagates in the quantum channel from the receiver 
to the sender of the signal photon. 
Recently, an entanglement distribution scheme 
against general collective noises with an efficiency proportional to $T$ 
has been proposed~\cite{2013Kumagai}. This scheme uses 
the above idea and another key idea proposed in Ref.~\cite{2005Yamamoto} 
which protects quantum states against general collective noises 
by using the two-photon DFS against the collective dephasing noise 
at the price of using two communication channels and receiving a constant loss. 
Such state protection is provided by 
the reciprocity of the quantum channel 
and a property of the quantum entanglement 
that disturbance on one half is equivalent to disturbance on the other half. 
In this paper, we report an experimental demonstration 
of the entanglement distribution scheme~\cite{2013Kumagai}
with an efficiency proportional to $T$ 
against collective noises including 
not only the phase noise but also the bit flip noise. 
We also show that the protection method is applicable
to distributing any single photon quantum state
with the use of the quantum parity check~\cite{2001Pittman}. 

\begin{figure}[t]
 \begin{center}
 \scalebox{0.53}{\includegraphics{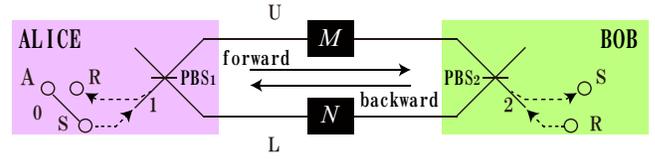}}
  \caption{(Color online) 
  Entanglement distribution protocol
  by using a counter-propagating photon as a reference light~\cite{2013Kumagai}. 
  When photons S and R are reached at port 2 and 1, respectively, 
  the entangled photon pair between Alice and Bob is extracted 
  by the quantum parity check on photons A and R. 
  \label{fig:scheme}}
 \end{center}
\end{figure}
We first review the protocol for sharing an entangled photon pair 
against general collective noises proposed in Ref.~\cite{2013Kumagai}, 
in which it is assumed that the party is allowed 
to use two noisy channels. 
The conceptual setup of the protocol is shown in Fig.~\ref{fig:scheme}. 
First the sender Alice prepares a maximally entangled photon pair 
A and S as 
$\ket{\phi^+}^{\A\S}
\equiv (\ket{\H}^{\A}\ket{\H}^{\S}+\ket{\V}^{\A}\ket{\V}^{\S})/\sqrt{2}$, 
and sends the signal photon S to Bob 
after injecting the photon to a polarization beamsplitter~($\U{PBS}\u{1}$) 
connected to two quantum channels from port 1. 
Here $\ket{\H}$ and $\ket{\V}$ represent 
horizontal~(H) and vertical~(V) polarization states 
of a photon, respectively. 
On the other hand, the receiver Bob prepares a reference photon R 
in the state $\ket{\D}^\U{R}\equiv (\ket{\H}^{\R}+\ket{\V}^{\R})/\sqrt{2}$, 
and sends it to Alice after injecting the photon 
to another PBS~($\U{PBS}\u{2}$) from port 2. 

At the noisy quantum channels,
unknown polarization transformations are added on the photons.
For the forward~(backward) propagation from Alice~(Bob) to Bob~(Alice), 
we denote the linear operators of 
lossy linear optical media 
for upper~(U) and lower~(L) optical paths in Fig.~\ref{fig:scheme}
by $M\u{f(b)}$ and $N\u{f(b)}$, respectively. 
The media satisfy 
$0 \leq M\u{f(b)}^\dagger M\u{f(b)} \leq 1$ and 
$0 \leq N\u{f(b)}^\dagger N\u{f(b)} \leq 1$. 
As is explained in Ref.~\cite{2013Kumagai}, 
for counter-propagating light pulses through any reciprocal media 
including the situation considered in this  paper, 
Alice and Bob can choose their coordinate systems 
such that the relationship between the action of the operators
on the single photons as 
\begin{eqnarray}
\bra{i}\Omega\u{b}\ket{j} = \bra{j}Z \Omega\u{f} Z \ket{i} 
\label{eq:Mfb}
\end{eqnarray}
are satisfied for $\Omega=M,N$ and $i,j\in \{ \H, \V \}$, 
where $Z=\ketbra{\H}{\H}-\ketbra{\V}{\V}$. 
In this paper, we chose the coordinate systems satisfying Eq.~(\ref{eq:Mfb}). 

After the transmission through the channels, 
the separated components of the photons S and R 
are recombined at $\U{PBS}\u{2}$ and $\U{PBS}\u{1}$, respectively. 
Alice and Bob postselect the events
of the photon S coming from port 2
and the photon R coming from port 1. 
As a result, the initial state $\ket{\phi^+}^{\A\S}\ket{\D}^\U{R}$
is transformed as 
\begin{widetext}
\begin{eqnarray}
 \ket{\phi^+}^{\A\S}\ket{\D}^\U{R}\rightarrow
  &(\bra{\H}{\Mf}\ket{\H}\bra{\H}{\Mb}\ket{\H}
\ket{\H}^\U{A}_0
\ket{\H}^\U{S}_2
\ket{\H}^\U{R}_1
+
\bra{\H}{\Mf}\ket{\H}\bra{\V}{\Nb}\ket{\V}
\ket{\H}^\U{A}_0
\ket{\H}^\U{S}_2
\ket{\V}^\U{R}_1\nonumber\\
&+
\bra{\V}{\Nf}\ket{\V}\bra{\H}{\Mb}\ket{\H}
\ket{\V}^\U{A}_0
\ket{\V}^\U{S}_2
\ket{\H}^\U{R}_1
+
\bra{\V}{\Nf}\ket{\V}\bra{\V}{\Nb}\ket{\V}
\ket{\V}^\U{A}_0
\ket{\V}^\U{S}_2
\ket{\V}^\U{R}_1)/2, 
\label{eq:MN}
\end{eqnarray}
\end{widetext}
where the subscripts represent the spatial ports of the photons. 
From Eq.~(\ref{eq:Mfb}), 
we see that the coefficients of the second and the third terms
in Eq.~(\ref{eq:MN}) are the same. 
Thus, 
by Alice performing the quantum parity check 
on the photons A and R, 
whose successful operation is described by 
$\ket{\H}\u{0}\bra{\H}\u{0}^\U{A}\bra{\V}\u{1}^\U{R}+
\ket{\V}\u{0}\bra{\V}\u{0}^\U{A}\bra{\H}\u{1}^\U{R}$, 
the maximally entangled state 
$
(\ket{\H}_0\ket{\H}^\U{S}_2+\ket{\V}_0\ket{\V}^\U{S}_2)/\sqrt{2}
$
is extracted. 
The success probability of the protocol is given by 
$|\bra{\H}{\Mf}\ket{\H}\bra{\V}{\Nf}\ket{\V}|^2/2$. 
In our experiment, 
we construct the lossy and noisy channels 
by $M\u{f(b)}=TU\u{f(b)}^\U{u}$ and $N\u{f(b)}=TU\u{f(b)}^\U{l}$, 
where $U\u{f(b)}^\U{u}$ and $U\u{f(b)}^\U{l}$ are unitary operators, 
and $T$ is a transmittance of an identical polarization-independent linear loss component. 
Since the two channels are independent, 
the success probability becomes $T^2 T\u{u}T\u{l}/2$, 
where $T\u{u}$ and $T\u{l}$ are given by 
average values of $|\bra{\H}{U\u{f}}^\U{u}\ket{\H}|^2$ 
about $U\u{f}^\U{u}$ and $|\bra{\V}{U\u{f}}^\U{l}\ket{\V}|^2$ about $U\u{f}^\U{l}$, respectively. 
If $U\u{f}^\U{u}$ and $U\u{f}^\U{l}$ are completely random, 
we obtain $T\u{u}=T\u{l}=1/2$. 
In the experiment, 
while we switch the unitary operators discretely as described later, 
the transmission is kept to $T\u{u}=T\u{l}=1/2$, 
resulting in the success probability of $T^2/8$. 

The efficiency $\mathcal{O}(T^2)$ of sharing the entangled states is improved 
by using a weak coherent light pulse~(wcp) instead of using the single photon for R. 
Suppose that average photon number of the wcp received by Alice is $\mu$, 
which means that Bob prepares the wcp of average photon number $\mu T^{-1}$. 
Since the quantum channel considered in this paper is a linear optical channel, 
the protocol with the use of the wcp for R works well as was described above 
when Alice receives one photon in the pulse R 
and Bob receives the signal photon S, 
which occurs at a probability of $\mathcal{O}(\mu T)$. 
Unfortunately, a conventional quantum parity check with linear optics 
and threshold photon detectors at Alice's side cannot perfectly 
discard the cases with multiple photons received in the pulse R. 
Such unwanted events occur at a probability of $\mathcal{O}(\mu^2 T)$ 
and they degrade the fidelity of the final state. 
As a result, by choosing the value of $\mu$ independent of $T$ 
such that $1\gg \mu$ is satisfied, 
a high-fidelity entangled photon pair is shared 
because of $\mathcal{O}(\mu T)\gg \mathcal{O}(\mu^2 T)$, 
with an overall success probability 
proportional to $T$ as $\mathcal{O}(\mu T)$. 

\begin{figure}[t]
 \begin{center}
 \scalebox{0.47}{\includegraphics{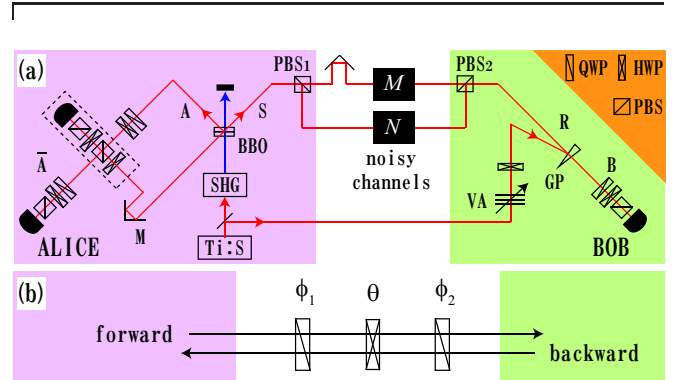}}
  \caption{(Color online) 
  (a)
  Our experimental setup.
  The component surrounded by a broken line is the quantum parity check
  for extracting and decoding the DFS.
  (b)
  The reciprocal noisy channel denoted by $M$ or $N$ in Fig.~\ref{fig:setup}~(a). 
  The operation is $Q(\phi_2)H(\theta)Q(\phi_1)$
  for the forward-propagating photons, and is
  $Q(-\phi_1)H(-\theta)Q(-\phi_2)$
  for the backward-propagating ones. 
  \label{fig:setup}}
 \end{center}
\end{figure}
The experimental setup is shown in Fig.~\ref{fig:setup}~(a). 
Light from 
a mode-locked Ti:sapphire~(Ti:S) laser~(wavelength: 790~nm; 
pulse width: 100~fs; repetition rate: 80~MHz) is divided into two beams. 
One beam is used at Alice's side. 
It is frequency doubled~(wavelength: 395~nm; power: 75~mW) 
by second harmonic generation~(SHG) 
for preparing $\ket{\phi^+}^{\A\S}$ 
through spontaneous parametric down conversion~(SPDC) 
by a pair of type-I and 1.5-mm-thick $\beta$-barium borate~(BBO) crystals.
The photon pair generation rate is 
$\gamma \approx 2\times 10^{-3}$. 
Photon S enters the two noisy channels after $\U{PBS}\u{1}$. 
After passing through the noisy channels, 
the H- and V-polarized components are extracted and recombined 
by $\U{PBS}\u{2}$, and goes to a photon detector in mode B 
after an wedged glass plate~(GP) whose reflectance is less than $10$\%. 
The other beam from the Ti:S laser is used 
to prepare a wcp as a reference light R at Bob's side. 
The intensity of the wcp is adjusted by a variable attenuator~(VA) 
in such a way that 
$\mu \approx 0.9\times 10^{-1}$ 
when it arrives at Alice's side after $\U{PBS}\u{1}$. 
The polarization of the wcp is set to the diagonal polarization
by a half wave plate~(HWP) after VA. 
The wcp R is reflected by GP and 
it propagates the noisy channels after $\U{PBS}\u{2}$ 
along the same spatial paths as photon S. 

After the transmission of photons through the quantum channels, 
Alice performs the quantum parity check 
for extracting the DFS and decoding to the qubit state of the single photon, 
which is shown in the broken box in Fig.~\ref{fig:setup}~(a). 
After the reference light pulse R passing through the HWP 
which flips H~(V) to V~(H), 
Alice mixes the light pulses A and R at a PBS 
with a temporal delay adjusted by mirrors~(M). 
Then she projects the photons
coming from one of the output ports of the PBS onto the diagonal polarization. 
Alice postselects the cases where 
at least one photons are detected at both output modes of the PBS. 
On the other hand, Bob postselects the cases 
where at least one photon appears in mode B. 
Under this post-selection rule, 
Alice and Bob share the photon pairs in modes $\Abar$ and B 
which are in state $\ket{\phi^+}\u{\Abar B}$ ideally. 
All detectors are silicon avalanche photon detectors 
which are coupled to single-mode optical fibers
after spectral filtering with a bandwidth of 2.7 nm. 

In this experiment,
we simulate a lossy depolarizing quantum channel for each noisy channel. 
The channel is composed of 
one HWP sandwiched by two quarter wave plates~(QWPs) as shown in Fig.~\ref{fig:setup}~(b). 
The operations of a HWP and a QWP acting on a single photon are described by 
$H(\theta)=\cos(2\theta) Z - \sin(2\theta) X$ 
and 
$Q(\phi)=(iI-\cos(2\phi) Z + \sin(2\phi) X)/\sqrt{2}$, respectively~\cite{2001James}, 
where
$I=\ketbra{H}{H}+\ketbra{V}{V}$, 
$X=\ketbra{H}{V}+\ketbra{V}{H}$ and $Y=-i\ketbra{H}{V}+i\ketbra{V}{H}$. 
Here $\theta$ and $\phi$ are rotation angles of the wave plates. 
The operation $Q(\phi_2)H(\theta)Q(\phi_1)$ 
on forward-propagating photons works as
$I$, $X$, $Y$ and $Z$ 
for the settings of 
$(\phi_1, \theta, \phi_2)=(0,0,0)$, 
$(0, -\pi/4, 0)$,
$(\pi/2, -\pi/4, 0)$ and
$(\pi/4, 0, \pi/4)$, 
respectively, up to global phases.
For the backward-propagating photons,
the operation of the channel becomes $Q(-\phi_1)H(-\theta)Q(-\phi_2)$. 
This results in the operation of the backward channel as 
$I$, $X$, $Y$ and $Z$ for the above four angle settings. 
For simulating the two depolarizing channels, 
and we slowly switched among the four settings 
of the wave plates independently in the two noisy channels. 
When we introduce the photon loss, 
we insert identical neutral density~(ND) filters in the two channels. 

We first performed the quantum state tomography~\cite{2001James} 
of the initial photon pair in modes A and S prepared by the SPDC. 
We reconstructed the density operator $\rho\u{AS}$ of the two-photon state 
with the use of the iterative maximum likelihood method~\cite{2007Rehacek}. 
The observed fidelity to the maximally entangled state $\ket{\phi^+}\u{AS}$ 
was $0.97\pm 0.01$, which shows 
Alice prepares a highly entangled photon pair. 

\begin{figure}[t]
 \begin{center}
  \scalebox{0.7}{\includegraphics{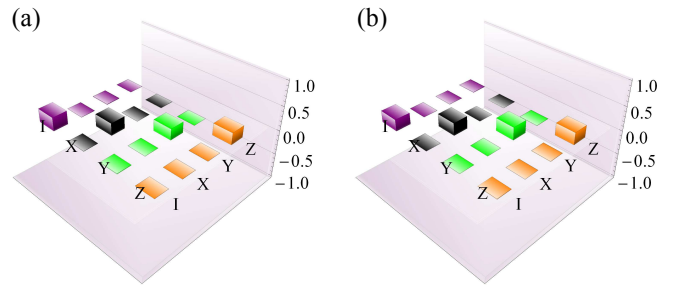}}
  \caption{(Color online) 
  (a) The real part of the reconstructed process matrix 
  for the forward propagation. 
  (b) is one for the backward propagation. 
  \label{fig:process}}
 \end{center}
\end{figure}
Before we perform the entanglement distribution by the DFS, 
we performed the process tomography~\cite{1997Chuang} of the noisy channel 
composed of the three wave plates 
for forward and backward propagation of photons. 
For this, we sent the photon S entangled with photon A to the noisy channel 
with the forward and backward configuration, 
and then we performed the state tomography of the two-photon output state. 
When we sequentially switch the angles of the wave plates 
for simulating $I$, $X$, $Y$ and $Z$, 
the process matrices of the channel are reconstructed 
as shown in Figs.~\ref{fig:process}~(b) and (c), 
by assuming that the initial state is 
the perfect entangled state $\ket{\phi^+}\u{AS}$. 
We see that the quantum channel well simulates 
the depolarizing channel for both directions. 

\begin{figure}[t]
 \begin{center}
  \scalebox{0.33}{\includegraphics{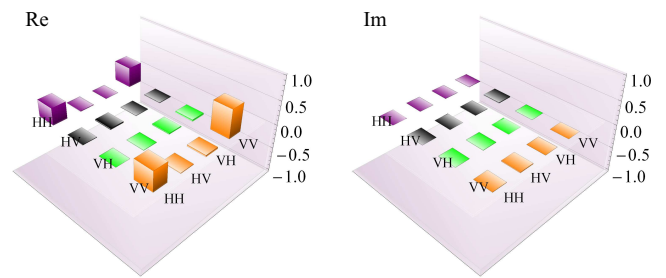}}
  \caption{(Color online) 
  The reconstructed density matrix for $T=1$. 
  \label{fig:matrixDFS}}
 \end{center}
\end{figure}
\begin{table}[t]
\begin{center}
\begin{tabular}
{c|ccc||c}
\hline
$T$  & Fidelity & Purity & EoF & $\U{Fidelity^\U{th}}$\\ \hline
 1  & $0.89\pm 0.02$ & $0.83\pm 0.03$ & $0.69\pm 0.06$ 
& $0.85\pm 0.01$\\
 0.48 & $0.85\pm 0.02$ & $0.79\pm 0.03$ & $0.63\pm 0.06$ 
& $0.83\pm 0.02$\\
 0.17 & $0.87\pm 0.02$ & $0.80\pm 0.04$ & $0.66\pm 0.06$ 
& $0.89^{+0.02}_{-0.03}$\\
\hline
\end{tabular}
 \caption{
 The experimental results 
 of the fidelity, purity and entanglement of formation~(EoF) of 
 the reconstructed density operators distributed by our scheme 
 for each transmittance $T$. 
 $\U{Fidelity^\U{th}}$ is the value predicted 
 by the theoretical model with the experimental parameters, 
 in which the error bars come from the observed values of the visibility 
 of the mode matching between the light pulses at Alice's side. 
 \label{tbl:view}}
 \end{center}
\end{table}
\begin{figure}[t]
 \begin{center}
  \scalebox{0.6}{\includegraphics{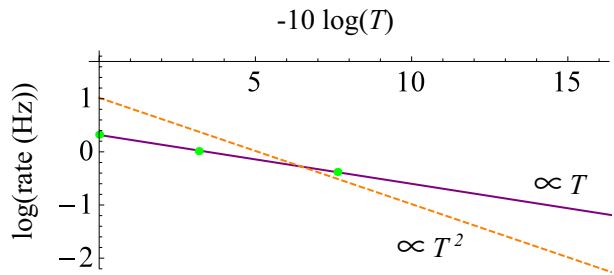}}
  \caption{(Color online) 
  The observed count rate of the states protected by the DFS. 
  The solid line fitted 
  to the experimental data proportional to $T^{0.92 \pm 0.02}$. 
  The broken line proportional to $T^2$ is expected 
  when the 2-qubit DFS method 
  with a forward-propagating reference single photon~\cite{2005Yamamoto} is used. 
  We assumed that the encoding and decoding are performed 
  by using a conventional linear optics~\cite{2005Yamamoto}. 
  The line passes through a value 
  $\mu^{-1}/2$ times as large as the observed rate for $T=1$. 
    \label{fig:result}}
 \end{center}
\end{figure}
Next, we performed our DFS scheme. 
When we inserted no ND filters to the channels
which is regarded as the case of transmittance $T=1$, 
the reconstructed density operator 
of the photon pair shared between Alice and Bob 
were shown in Fig.~\ref{fig:matrixDFS}. 
The fidelity, the purity and entanglement of formation~(EoF) 
of $\rho\u{AB}^{T=1}$ are $0.89\pm 0.02$, $0.83\pm 0.03$ and $0.69\pm 0.06$, respectively. 
The result shows that the DFS scheme protects 
the entanglement against collective depolarizing noises. 
When we inserted ND filters to the channels 
for $T$ to be $\approx 0.48$ and $\approx 0.17$, 
in order for $\mu$ to be a constant at Alice's side, 
we chose the intensity of the reference light pulse R 
at Bob's side to be $T^{-1}$ times as high as that for the case of $T=1$. 
The observed fidelity, the purity and EoF of the reconstructed state 
for each transmittance are shown in Table~\ref{tbl:view}. 
For all $T$, the highly entangled photon pairs were shared between Alice and Bob. 
The sharing rate of the final states for each $T$ is shown in Fig.~\ref{fig:result}, 
which shows that the sharing rate is proportional to $T$. 
Fidelities predicted by experimental parameters 
with a theoretical model 
which deals with multiple photon emission events and 
mode matching between the photon in mode A and the wcp 
are shown in Table~\ref{tbl:view}, 
and they are in good agreement with the observed values. 
In the model, 
the main causes of the degradation of the entanglement 
are multiple photons from the photon pair and the coherent light pulse, 
and mode mismatch at the quantum parity check. 
The ratios of both errors 
to the desired events are almost independent of the transmittance $T$. 
As a result, the DFS scheme will work well as long as 
the count rate is much larger than the dark count rate of the detectors. 

\begin{figure}[t]
 \begin{center}
 \scalebox{0.6}{\includegraphics{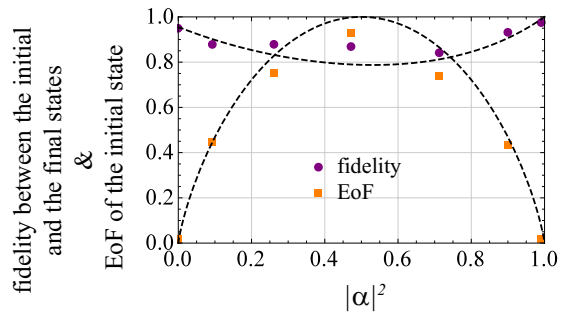}}
  \caption{(Color online)
  The fidelity between the initial and 
  the final states~(circle),
  and the EoF of the initial state~(square)
  for various values of $|\alpha|^2$ 
  which is estimated from the probabilities 
  of the component $\ket{\HH}$ of the initial state. 
  A dashed curve for EoF is given 
  by using the ideal state of $\ket{\phi_{\alpha,\beta}}$. 
  A dashed curve for the fidelity is 
  given by the inner product between 
  $\ket{\phi_{\alpha,\beta}}$ and the final state 
  with considering the multiple emission events 
  and the mode matching effect. 
  \label{fig:comparison}}
 \end{center}
\end{figure}
The DFS scheme is applicable to the protection of 
not only the maximally entangled state but also a state 
in the form of $\ket{\phi_{\alpha,\beta}}=\alpha\ket{\HH}+\beta\ket{\VV}$, 
where $\alpha$ and $\beta$ are arbitrary complex numbers 
satisfying $|\alpha|^2+|\beta|^2=1$. 
To see this,
we prepared such non-maximally entangled states 
by rotating the polarization of the pump light, 
and then performed the DFS method. 
We reconstructed the initial and the final states, 
and calculated the fidelity between the two states. 
From the experimental result shown in Fig.~\ref{fig:comparison}, 
the fidelity between the initial and the final states
is larger than $0.84$ for any value of $|\alpha|^2$. 
As a result, we see that our DFS method is useful 
for sharing the state $\ket{\phi_{\alpha,\beta}}$. 
By using the parity check, 
any state in $\alpha\ket{\H}+\beta\ket{\V}$ can be encoded 
to the form of $\ket{\phi_{\alpha,\beta}}$ 
without revealing the values of $\alpha$ and $\beta$. 
Decoding can be performed by measuring the photon at Alice's side
by $\ket{\D}$ after the DFS scheme. 
This means that any single qubit can be sent
from Alice to Bob by using the DFS scheme. 

We note that the DFS scheme uses two interferometers,
one of which is for sending the photons by two quantum channels 
and the other is for performing the quantum parity check at Alice's side. 
In the former interferometer, 
while unknown phase shift between H and V 
may be added by the fluctuations of the two quantum channels, 
it is automatically canceled by the DFS. 
Thus, it is insensitive to the timing mismatch 
between the photons passing through the two arms. 
The latter interferometer used for the quantum parity check 
is also insensitive to the timing mismatch 
between the photons from the SPDC and the coherent light pulse R 
because the two-photon interference~\cite{1987Hong} is used.
In fact, the observed FWHM of the visibility against the timing mismatch
is $\sim$ 150 $\mu$m which is much longer than the wavelength of the photons. 
As a result, the DFS method is totally robust 
against the fluctuations of the optical circuit, 
and does not need the control with the wavelength-order precision. 

In conclusion, we have demonstrated 
the robust entanglement-sharing scheme 
over collective noisy channels.
By using the counter-propagating reference classical light 
with an intensity 
inversely proportional to the transmittance $T$ of the quantum channel, 
we experimentally achieved the entanglement-sharing rate proportional to $T$. 
We also demonstrated that the DFS method is used 
to distribute any state in $\alpha \ket{\HH}+\beta\ket{\VV}$, 
which indicates that it is applicable to distributing any unknown single qubit 
with the redundant encoding by the quantum parity check. 
The essence of the scheme is 
the use of the reciprocal property of the channel 
and the property of the entanglement 
that a disturbance of one half is equivalent to that of the other half.
Since optical fibers are known to be reciprocal media 
and its 
fluctuations are slow enough to satisfy the collective assumption, 
we believe that our efficient DFS method will be useful to distribute entanglement 
for optical fiber communication over a long distance. 
In addition, 
it may open up a new sensor 
for detecting non-reciprocal property of the noisy channel 
by measuring the quantity of entanglement. 

We thank Hidetoshi Kumagai for helpful discussion. 
This work was supported by 
MEXT/JSPS KAKENHI Grant Number 
JP26286068, 
JP15H03704, 
JP16H02214, 
JP16K17772.


\begin{thebibliography}{10}

\bibitem{1984Bennett}
C.~H. Bennett and G.~Brassard,
\newblock Proceedings of IEEE International Conference on Computers, Systems,
  and Signal Processing, Bangalore, India , 175 (1991).

\bibitem{1991Ekert}
A.~K. Ekert,
\newblock Phys. Rev. Lett. {\bf 67}, 661 (1991).

\bibitem{1992Bennett}
C.~H. Bennett, G.~Brassard, and N.~D. Mermin,
\newblock Phys. Rev. Lett. {\bf 68}, 557 (1992).

\bibitem{2011Sangouard}
N.~Sangouard, C.~Simon, H.~De~Riedmatten, and N.~Gisin,
\newblock Reviews of Modern Physics {\bf 83}, 33 (2011).

\bibitem{2009Broadbent}
A.~Broadbent, J.~Fitzsimons, and E.~Kashefi,
\newblock Universal blind quantum computation,
\newblock in {\em Foundations of Computer Science, 2009. FOCS'09. 50th Annual
  IEEE Symposium on}, pp. 517--526, IEEE, 2009.

\bibitem{2016Takeuchi}
Y.~Takeuchi, K.~Fujii, R.~Ikuta, T.~Yamamoto, and N.~Imoto,
\newblock Phys. Rev. A {\bf 93}, 052307 (2016).

\bibitem{2000Kwiat}
P.~G. Kwiat, A.~J. Berglund, J.~B. Altepeter, and A.~G. White,
\newblock Science {\bf 290}, 498 (2000).

\bibitem{2003Walton}
Z.~D. Walton, A.~F. Abouraddy, A.~V. Sergienko, B.~E.~A. Saleh, and M.~C.
  Teich,
\newblock Phys. Rev. Lett. {\bf 91}, 087901 (2003).

\bibitem{2004Boileau}
J.-C. Boileau, D.~Gottesman, R.~Laflamme, D.~Poulin, and R.~W. Spekkens,
\newblock Phys. Rev. Lett. {\bf 92}, 017901 (2004).

\bibitem{2004Bourennane}
M.~Bourennane {\em et~al.},
\newblock Phys. Rev. Lett. {\bf 92}, 107901 (2004).

\bibitem{2004Boileau2}
J.-C. Boileau, R.~Laflamme, M.~Laforest, and C.~R. Myers,
\newblock Phys. Rev. Lett. {\bf 93}, 220501 (2004).

\bibitem{2005Yamamoto}
T.~Yamamoto, J.~Shimamura, {\mbox{\c{S}}}.~K. \"Ozdemir, M.~Koashi, and
  N.~Imoto,
\newblock Phys. Rev. Lett. {\bf 95}, 040503 (2005).

\bibitem{2006Chen}
T.-Y. Chen {\em et~al.},
\newblock Phys. Rev. Lett. {\bf 96}, 150504 (2006).

\bibitem{2007Prevedel}
R.~Prevedel {\em et~al.},
\newblock Phys. Rev. Lett. {\bf 99}, 250503 (2007).

\bibitem{2007Yamamoto}
T.~Yamamoto {\em et~al.},
\newblock New Journal of Physics {\bf 9}, 191 (2007).

\bibitem{2008Yamamoto}
T.~Yamamoto, K.~Hayashi, {\c{S}}.~K. {\"O}zdemir, M.~Koashi, and N.~Imoto,
\newblock Nature Photonics {\bf 2}, 488 (2008).

\bibitem{2003Lidar}
D.~A. Lidar and K.~B. Whaley,
\newblock Decoherence-free subspaces and subsystems,
\newblock in {\em Irreversible Quantum Dynamics}, pp. 83--120, Springer, 2003.

\bibitem{2001Kempe}
J.~Kempe, D.~Bacon, D.~A. Lidar, and K.~B. Whaley,
\newblock Phys. Rev. A {\bf 63}, 042307 (2001).

\bibitem{2011Ikuta}
R.~Ikuta {\em et~al.},
\newblock Phys. Rev. Lett. {\bf 106}, 110503 (2011).

\bibitem{2013Kumagai}
H.~Kumagai, T.~Yamamoto, M.~Koashi, and N.~Imoto,
\newblock Phys. Rev. A {\bf 87}, 052325 (2013).

\bibitem{2001Pittman}
T.~B. Pittman, B.~C. Jacobs, and J.~D. Franson,
\newblock Phys. Rev. A {\bf 64}, 062311 (2001).

\bibitem{2001James}
D.~F.~V. James, P.~G. Kwiat, W.~J. Munro, and A.~G. White,
\newblock Phys. Rev. A {\bf 64}, 052312 (2001).

\bibitem{2007Rehacek}
J.~$\U{\check{R}eh\acute{a}\check{c}ek}$, Z.~Hradil, E.~Knill, and A.~I.
  Lvovsky,
\newblock Phys. Rev. A {\bf 75}, 042108 (2007).

\bibitem{1997Chuang}
I.~L. Chuang and M.~A. Nielsen,
\newblock Journal of Modern Optics {\bf 44}, 2455 (1997).

\bibitem{1987Hong}
C.~Hong, Z.~Ou, and L.~Mandel,
\newblock Physical Review Letters {\bf 59}, 2044 (1987).

\end{thebibliography}
\end{document}